\documentclass[numberedappendix]{emulateapj}
\usepackage[]{amsmath}
\usepackage{xcolor}
\usepackage{graphicx}
\usepackage{multirow}

\def\um{\mu\textrm{m}}

\def\cm1{\textrm{cm}^{-1}}

\def\shat0{\hat{\textbf{s}}_0}
\def\Ds{\xi}

\def\samp{s}
\def\baseline{\mathbf{B}}
\def\B{B}
\def\F{\mathcal{F}}
\def\I{\textit{I}}
\def\Im{\I_\textrm{measured}}
\def\mI{\mathcal{I}}
\def\xn{x_n}
\def\xnp{x_{n'}}
\def\pxn{(\xn)}
\def\pxnp{(\xnp)}
\def\Ixn{\mI\pxn}
\def\Ikxn{\mI_k\pxn}
\def\Iksxn{\mI_k^{*}\pxnp}
\def\Vb{\mathcal{V}_\baseline}
\def\Phib{\Phi_\mathbf{B}}

\def\Vi{\mathcal{V}_\textrm{i}}
\def\Phii{\Phi_\mathbf{i}}
\def\V{\mathcal{V}}
\def\S{\mathcal{S}}
\def\R{\mathcal{R}}
\def\oS{\overline{\S}}

\def\etamf{\eta_\textrm{mf}}

\def\s{\sigma}
\def\sig{\Delta}

\def\Phir{\Phi_r}
\def\etaD{\eta_D}
\def\Tbp{\mathcal{T}_\textrm{bp}}
\def\A{\mathcal{A}}
\def\F{\mathcal{F}}
\def\Ahat{\hat{\mathcal{A}}}
\def\Dx{dx}
\def\Dt{dt}
\def\D{d}

\def\sinc{\textrm{sinc}}
\def\real{\textrm{Re}}
\def\imag{\textrm{Im}}

\def\DFT{\mathbf{DFT}}
\def\dsig{\delta\s}
\def\varPhir{\sig^2_\Phi}

\def\varopd{\sig^2_{\textrm{OPD}}}
\def\wk{w_k}

\def\SNR{\textrm{SNR}}

\def\ni{n_\mI}
\def\nA{n_+}
\def\nB{n_-}
\def\NEP{\textrm{NEP}}
\def\NEPtot{\NEP_\textrm{tot}}
\def\NEPph{\NEP_\textrm{ph}}
\def\NEPdet{\NEP_\textrm{det}}
\def\NEPsou{\NEP_\textrm{sou}}
\def\VAR{\mathbf{VAR}}
\def\varI{\sig_\mI^2}
\def\sigI{\sig_\mI}
\def\varspec{\sig_\mathcal{S}^2}

\def\sigspec{\sig_\mathcal{S}}

\def\sigalpha{\sig_\alpha}
\def\sigL{\sig_L}

\def\inst{\textrm{inst}}
\def\col{\textrm{col}}

\begin{document}

\title{Far-Infrared double-Fourier interferometers and their spectral sensitivity}
\author{Maxime J. Rizzo, Lee G. Mundy, Arnab Dhabal,  Dale J. Fixsen}

\affil{Department of Astronomy, University of Maryland, College Park, MD 20742}
\author{Stephen A. Rinehart, Dominic J. Benford, David Leisawitz, Robert Silverberg, Todd Veach}
\affil{NASA Goddard Space Flight Center, Greenbelt, MD 20771}

\author{Roser Juanola-Parramon}
\affil{University College  London, London, UK}
\email{mrizzo@astro.umd.edu}

\begin{abstract}
Double-Fourier interferometry is the most viable path to sub-arcsecond spatial resolution for future 
astronomical instruments that will observe the universe at far-infrared wavelengths. The double transform spatio-spectral
interferometry couples pupil plane beam combination with detector arrays to enable imaging spectroscopy of
wide fields, that will be key to accomplishing top-level science goals.
The wide field of view and the necessity for these instruments to fly above the opaque atmosphere create unique characteristics and requirements compared to instruments on ground-based
telescopes.
In this paper, we discuss some characteristics of single-baseline spatio-spectral interferometers. We investigate the impact of intensity and optical path difference noise on the interferogram and the spectral signal-to-noise ratio. We apply our findings to the special case of the Balloon Experimental Twin Telescope for Infrared Interferometry (BETTII), a balloon payload that will be a first application of this technique
at far-infrared wavelengths on a flying platform. 
\end{abstract}

%


\section{Introduction}
Observations at mid- to far-infrared wavelengths from the Earth's surface are extremely 
limited by the large atmospheric opacity in this region of the spectrum. Space-based telescopes 
like IRAS \cite[12-100 $\um$;][]{1984ApJ...278L...1N}, ISO \cite[2.5-240 $\um$;][]{1996A&A...315L..27K}, \textit{Spitzer} \cite[3.6-160 $\um$;][]{2004ApJS..154....1W}, AKARI  \cite[1.7-180 $\um$;][]{2007PASJ...59S.369M}, WISE \cite[3.4-22 $\um$;][]{2010AJ....140.1868W} and \textit{Herschel} \cite[55-672 $\um$;][]{2010A&A...518L...1P} have demonstrated the scientific value of observations at 
these wavelengths; but the spatial resolution of space-based observatories is limited by the cost 
and complexity of building and flying progressively larger aperture telescopes. 
Interferometry is a common solution to this problem on the ground, and is a viable path forward to obtain much
higher resolution than what single apertures can provide. 
In particular, spatio-spectral interferometry \citep{Mariotti:1988vea} is a way to achieve 
high angular and spectral resolutions at far-IR wavelengths from above the atmosphere, without the cost and limitations of large single apertures. 

Several space-based interferometer concepts, the Far Infrared Interferometer \citep[FIRI;][]{2009ExA....23..245H}, the Space Infrared Interferometer Telescope
\citep[SPIRIT;][]{Leisawitz:2007if}, and the Submillimeter Probe of the Evolution of Cosmic Structure \citep[SPECS;][]{Harwit:2006hl}, have been proposed and use spatio-spectral interferometry to achieve the much needed angular resolution to 
study astronomical processes such as the birth of stars and planetary systems, the activity in 
galactic nuclei and the formation of galaxies in the distant universe. The FIRI and SPIRIT concepts have
two mirrors which are movable on one axis along a monolithic truss to provide a range of baseline lengths.
SPECS consists of three spacecraft connected via tether to achieve baselines of order 1~km. 

There are numerous engineering challenges to be addressed before such missions can become reality. A number of them can be tackled with testbeds \citep[e.g.][]{Leisawitz:2012ik, 2012ApOpt..51.2202G} and small-scale pathfinder missions. These missions 
will likely be two-element, single baseline interferometers in space or on balloon platforms,
such as the Balloon Experimental Twin Telescopes for Infrared Interferometry \cite[BETTII;][]{2014PASP..126..660R} and to a certain extent the Far-Infrared Interferometric Telescope Experiment \cite[FITE;][]{2010TrSpT...7.Tm47K}.  
These pathfinders will have very limited baseline coverage and
rather than producing full images, they will focus on reconstructing 
spectral information from closely-spaced sources. This paper explores
aspects of the noise in spectral measurements specific to these instruments.

\subsection{Spatio-spectral interferometry}

In their pioneering paper, \cite{Mariotti:1988vea} lay out the principles of spatio-spectral 
(or double-Fourier) interferometry. A spatio-spectral interferometer consists of a Fourier transform 
spectrometer (FTS), where a delay line mechanism modulates the optical path difference (OPD) between two independent light beams before combining them in the pupil plane. The instrument produces interferograms, which are arrays of power measurements as a function of the OPD. Unlike traditional FTS, 
where a single incoming beam is split, delay-modulated, and recombined, a double-Fourier 
interferometer utilizes multiple light collectors pointing to the same astronomical source and 
combines the incoming light from the collectors pairwise in the pupil plane. The orientation 
and magnitude of the baselines - the vectors between each pair of light collectors - determines 
which  spatial frequency of the astronomical image the instrument measures. 
Longer baselines correspond to higher angular resolutions. The ``double-Fourier" aspect comes from 
the fact that the interferogram measured on a given baseline is related to the Fourier Transform (FT)
of the spatial and spectral distribution of the source emission.
Two FTs are used to reconstruct the full spatio-spectral datacube representing the 
astronomical scene: the spectra which are more directly related to the power as a function of time
delay difference between the two incoming beams (equivalent to the OPD) and
the source 2D spatial structure on the sky which is more directly related to measurements
accumulated from many different baseline vectors. The length of the baseline vectors can be changed by modifying the distance between the light collectors. The orientation of the vectors can be changed by rotating the baseline with respect to the source on the sky.
The plane representing the source's interferometric visibilities
as a function of baseline vector is referred to as the ($u, v$)-plane and is a common notion in ground-based submillimeter and radio interferometry. This paper focuses on the reconstruction of the spectrum from closely-spaced point sources using single-baseline measurements, and does not address the techniques and sensitivities involved in using multiple baseline lengths to produce an image of the scene; a mathematical formalism that covers imaging is already proposed in \cite{Elias:2007jsa}.


Proposed double-Fourier instruments at far-IR wavelengths distinguish themselves from operating interferometers at sub-millimeter and radio wavelengths in several ways. First, they do not directly measure the phase information. The fundamental measurement is a time series of real-valued power as a delay line modulates the OPD in a controlled sequence (for example a linear ramp). The OPD from the delay line, as well as other OPD contributors in each arm of the instrument, and the external OPD created when the line of sight to a source is not perpendicular to the baseline vector, add up to the total OPD. For a given detector location along the projected baseline vector, there exists a value of the OPD in the delay line that exactly compensates all other OPD contributors. This delay line position results in a zero net total OPD, and is called the Zero Path Difference (ZPD). At this value of OPD, an incoming plane wave traverses the two beam paths reaching the detector exactly with the same phase, for all wavelengths. ZPD corresponds to the center of an interferogram for a source at that detector location. In the
context of this paper, the phase for a given wavelength $
\phi_\lambda$ is related to the OPD between the beams from each arm when they combine, at the time of a data point measurement: $\phi_\lambda = 2\pi\textrm{OPD}/ \lambda$.



A second important difference for balloon and space interferometers is that collectors are not fixed
to the Earth. In the case of BETTII and SPIRIT, the collectors are fixed to a truss structure which
is part of the mechanical system for pointing the collectors. Consequently, baseline length
and external OPD, as relevant to an astronomical source, are not independent of pointing errors. The impact of errors in baseline length is modest because the relevant measure is in terms of fractions of the collector diameter. Errors in pointing translate into external OPD as the sine of the error angle times the baseline length, while the relevant measure is the wavelength. This can easily become significant;
for example, a 1" pointing error for an 8~m long baseline corresponds to a $38~\um$ shift in OPD.

\begin{figure}[ht!]
\begin{center}
\plotone{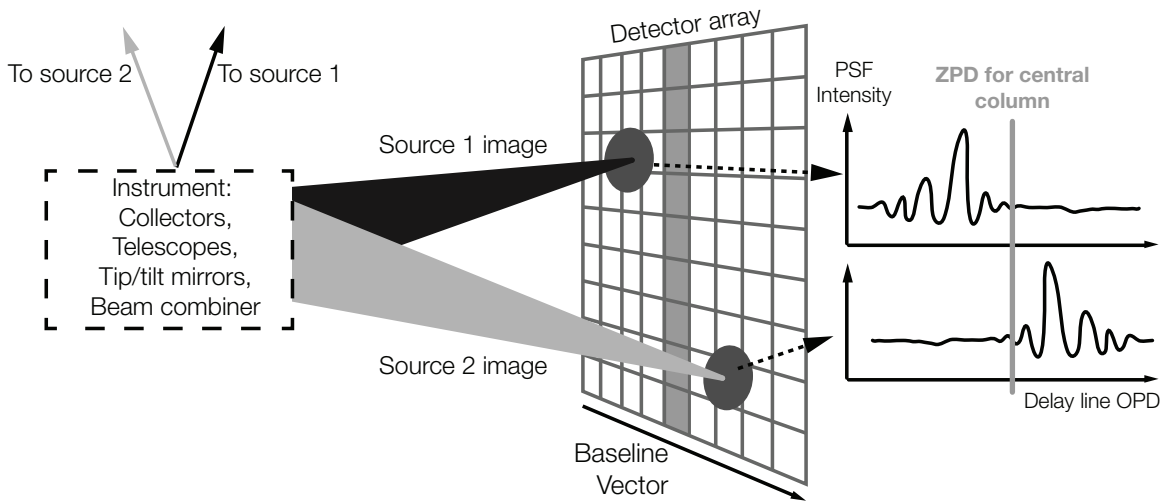}
\caption[WideField]{Concept of wide-field double-Fourier interferometry. Light from the instrument is focused after combination to an image of the sky on the detector array (represented as the grid). Each column of the detector has a distinct ZPD so the interferometric responses (right side) of two sources on different columns are centered around different delay positions. The gray stripe represents the central column on the detector array and its corresponding ZPD on the interferograms.}
\label{fig:widefield}
\end{center}
\end{figure}

Third, bolometer-type detectors, such as being built for BETTII and envisioned for SPIRIT, are
easily, and indeed typically, configured as two-dimensional arrays. With pupil plane combination,
the entire field of view has an interferometric response; hence wide-field interferometry over
multi-pixel arrays is straightforward. Fig.~\ref{fig:widefield} shows this concept and sketches the instrumental
response. For the configuration shown with the detector array columns aligned perpendicular
to the baseline vector, ZPD is the same along lines perpendicular to the baseline vector projected on the detector. As the OPD is swept, it moves across ZPD for the different columns in the array, yielding interferograms with shifted centers corresponding to the changes in external OPD for each source location in the field. 

By sweeping the OPD, the double-Fourier instrument measures interferograms which contain both spectral and spatial information over the detector array. The full spatial and spectral source information can be unambiguously recovered by repeating the delay line sweep over a range of baseline angles and lengths, which correspond to different spatial frequencies on the sky \citep{Mariotti:1988vea}.

\subsection{The case study: BETTII}

The BETTII project \citep{2014PASP..126..660R}, is a motivation for this paper and a near-term application of spatio-spectral interferometry. BETTII consists of two 50~cm siderostats on a fixed 8~m baseline, with a far-IR beam-combining instrument at the center. It will observe the far-IR universe in two 
wavelength bands, 30-55~$\um$ (``band 1") and 60-90~$\um$ (``band 2"). The instrument is currently under construction at NASA Goddard Space Flight Center and is scheduled to launch in the Fall of 2016 on a stratospheric balloon from Fort Sumner, New Mexico, to an altitude of 35~km in order to be above most of the atmosphere. For its first flight, BETTII will focus on the study of dense star formation in nearby clusters. While a complete image reconstruction is not possible due to the static baseline length, BETTII will help resolve point source objects that are 0.5-1" apart in band 1 and 2, respectively, more than ten times the spatial resolution of \textit{Spitzer} at 24~$\um$ and six times the resolution of SOFIA at 37~$\um$.  Combined with a modest spectral resolution of $\R=10 - 50$, BETTII will measure the spectral energy distributions (SEDs) of clustered young stars to determine their evolutionary stage, locate the origin of the far-IR emission, and improve our understanding of how stars accrete their mass in these very dense regions of stellar birth \cite[e.g. see][and references therein]{2014prpl.conf..149T}. For resolved sources, the fixed baseline will not completely lift degeneracies between the spectral and spatial information; however detailed source modeling can put constraints on the distribution of the far-IR emission
\citep[e.g][]{2013ApJS..207...30W}.

In this paper, we study how various types of noise propagate to the derived spectrum in an
instrument like BETTII or SPIRIT. In section~\ref{sec:formalism}, we establish a mathematical formalism that can be used to represent interferograms. In section~3, we look at the dominant types of noise in the interferogram and define the relevant timescales associated with spatio-spectral interferometers. In section~4, we derive the spectral signal-to-noise ratio ($\SNR$). In section~5, we apply these results to the special case of BETTII to derive its point source spectral sensitivity.

\section{Mathematical formalism}
\label{sec:formalism}
\begin{figure}[ht!]
\begin{center}
\plotone{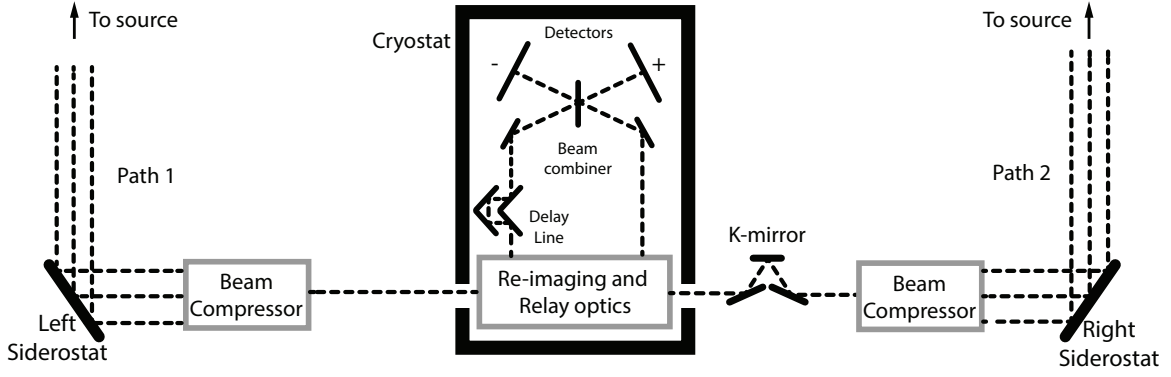}
\caption[optics]{Optical train diagram of a typical far-IR, double-Fourier instrument. The K-mirror rotates the beam to align the fields of view of the two sides. Inside the cryostat, a set of optics re-image the pupil, implement a controlled instrumental delay between them with the Delay Line, and relay them towards the central beam combiner. After the combiner, the beams are imaged onto the detectors. To see the BETTII-specific implementation of this design, see \cite{2014PASP..126..660R}.}
\label{fig:optics}
\end{center}
\end{figure}

The general optics layout for a double-Fourier system is shown in Fig.~\ref{fig:optics} for a single baseline. 
The combination of the siderostat and beam compressor acts as an afocal telescope 
which outputs a  parallel beam with a diameter convenient for the rest of the optical train.
The K-mirror in one beam path corrects for the pupil rotation so that the
images of the sky from the two collectors are matched over the field of view.
At the center of the instrument, there are optics for pupil re-imaging, filtering, and beam folding, as required by the specific implementation.
The key components for our purpose are the delay line, beam combiner and detectors. 
The delay line introduces a controlled OPD between both arms.
The two incoming beams are combined in the outputs from the beam combiner. 
We arbitrarily define one output as the ``+" and the other as the \mbox{``-"}. 
To conserve photon energy, the two outputs must be complimentary such that the summed power of the two is
independent of the OPD. In an ideal double-Fourier system, the two beam paths are symmetric about ZPD; 
hence, the power 
from the ``+" and ``-" outputs are equal at ZPD, and have odd symmetry about ZPD. In a traditional FTS at ZPD, one output has fully constructive interference while the other has fully destructive interference, with even symmetry about ZPD.

\subsection{Interferograms for a two-aperture instrument}

The interferogram for a single frequency of light measured at the outputs of the ideal, two-element double-Fourier instrument can be 
described in terms of the normalized intensity:
\begin{equation}
\hat{I}_\pm(x,\s) = \real(1 \pm i \; \Vb(\s) e^{-2i\pi \s x}),
\label{eq:basicinterferogram}
\end{equation}
where $\s \equiv {1 \over \lambda}$ is the wavenumber of the light in $\cm1$ as per the convention for the FTS literature, $x$ is the instrumental OPD created by the delay line with $x=0$ corresponding to ZPD, and $\Vb(\s)$ is the complex spatial visibility
 of the astronomical source for the baseline vector $\baseline$. ``$\real(f)$" indicates the real part of the complex-valued function $f$. The~$\pm$ indicates values for the two output beams: ``+" and ``-" in Fig.~\ref{fig:optics}.
The derivation of this expression is given in Appendix~\ref{ap:interfero}.

The normalized complex spatial visibility $\Vb$ has a magnitude of 1 for all baselines for which the source is completely unresolved. For extended sources, the spatial visibility depends on the source geometry, intensity distribution, and the instrument baseline vector as described in Chapter~2 of \cite{2000plbs.conf.....L} 
and Chapter~3 of \cite{Thompson:2008ww}.  
For a normalized source brightness distribution $\hat{\F}$, the spatial visibility with respect to a phase reference position on the sky can be written as:
\begin{equation}
\Vb(\s)  =  \int_\textrm{source} d\Omega \Ahat(\Ds) \hat{\F}(\Ds ) e^{-2i\pi\s\Ds\cdot \baseline},
\label{eq:viseq}
\end{equation}
where $\Ahat$ is the normalized reception pattern of the collecting area; $\baseline$ is the baseline vector between the two collectors and $\Ds$ is the vector on the plane of the sky from the phase reference position to the infinitesimal solid angle $d\Omega$. The resulting visibility as a function of baseline vector is the 2-dimensional FT of the source's sky distribution. 
Since $\hat{\F}$ does not have to be symmetric with respect to the chosen phase center, $\Vb$ is in general complex and can be expressed as an amplitude and a phase, $\Phib(\s)$: $\Vb(\s)  =  |\Vb(\s)|e^{i\Phib(\s)} $.


Real instruments have asymmetries, imperfections, and measurement errors which can create phase-shifts between the two optical paths and across the pupils.
Fixed instrumental effects 
can be represented by a normalized instrumental visibility loss term, 
$\Vi(\s)$ where the complex quantity $\Vi(\s) = |\Vi(\s)|e^{i\Phii(\s)}$, as described in details in Chapter~3 of \cite{2000plbs.conf.....L}, represents both amplitude losses and phase shifts (see Appendix~\ref{ap:interfero}). Additional phase errors can arise from
imperfect knowledge of the real-time optical path lengths which we will represent as
$e^{i\Phir(\s, x)}$, where $\Phir(\s, x)$ is the ``phase noise"; this term depends on the OPD $x$ through time-dependent phenomena such as mechanical jitters, temperature variations in
the optics support, or pointing errors. In the rest of this paper, we will mostly talk about this ``OPD noise", which is the physical source of the noise, whereas phase noise represents its effects on the interferogram.
The total complex visibility sampled at a single $\s$ by the system is $\Vb(\s)\Vi(\s) e^{i\Phir(\s, x)}$, and it is normalized such that, for an ideal instrument observing a point source, this quantity is equal to 1 at ZPD.

Using Eq. \ref{eq:basicinterferogram} for the monochromatic source, the polychromatic interferogram is the integral over $\s$ of this dimensionless response at each wavenumber. The total amount of power coming into the 2-aperture interferometer within a small wavenumber range $d\s$ is $2\A \B(\s) c d\s$ where $2\A$ is the total aperture area in m$^2$, $\B(\s)$ is the spectral flux density in W$\cdot$m$^{-2}\cdot$Hz$^{-1}$ and $c$ is the speed of light in cm$\cdot$s$^{-1}$. 
Filters and optics in an instrument cause a wavenumber-dependent transmission profile $\Tbp(\s)$. The quantum efficiency of the detector can depend on wavenumber, $\etaD(\s)$. For multi-pixel detectors the interferogram is measured by matched filtering a point-spread function on a pixel array, which has some efficiency $\etamf$.


The actual power measured by the instrument can be represented as:
\begin{align}
I_\pm(x) &= \A c\int_0^{+\infty} \etamf\etaD\Tbp \B \times\nonumber\\
& \quad  \real \left[\left(1\pm i \Vi\Vb e^{i\Phir} e^{-2i\pi \s x}\right) \right]d\s,
\end{align}
where the factor of 2 for the two apertures is dropped because it is implicit in Eq. \ref{eq:basicinterferogram}. All quantities within the integral can be functions of wavenumber. All the static interferometric loss terms and delay-dependent phase errors are in $\Vi$ and $e^{i\Phir}$, respectively.

Instead of considering each separate output, we use $\I = \I_+ - \I_-$ as our interferogram expression, which cancels out the constant term. We also introduce an interferometric instrument transmission function, which can be complex, which represents the normalized amplitude and phase of the interferogram for a point source of uniform spectrum and no phase noise:
\begin{equation}
T_{\inst}(\s) \equiv \A c\etamf\etaD\Tbp \Vi = |T_{\inst}(\s)|e^{i\Phi_{\inst}(\s)},
\end{equation}

 We can then write the modulated signal as:
\begin{equation}
I(x) = \real\left( 2\int_{0}^{+\infty} i |T_{\inst}| \B \Vb e^{i\Phir + i\Phi_{\inst}} e^{-2i\pi \s x}d\s \right),
\label{eq:modsignal}
\end{equation}
where $\B$ is real and $\Vb$ can be complex.

Eq.~\ref{eq:modsignal} can be turned into a Fourier transform by mirroring all quantities to negative wavenumbers. This
convention is explained in detail in \citet{Davis:2001tr} for FTS instruments; the odd symmetry of the interferogram for a system with one beam combiner
and the complex instrumental transfer function means that the incident spectrum on the detectors
must be mirrored to -$\s$ as the negative of the complex conjugate of +$\s$: 
$\S_e(\s) \equiv [T_{\inst} \B \Vb]_e(\s) = {1 \over 2}\left[T_{\inst}(\s) \B(\s) \Vb(\s) - T^*_{\inst}(-\s) \B(-\s) \Vb^*(-\s)\right]$. 
We use the subscript ``$e$" to denote the reflected function, and will apply this convention in the rest of this paper; this reflection ensures that the integrals keep the same value when are expressed from $-\infty$ to $+\infty$, and does not affect the $\SNR$ estimates: although the signal appears to be divided by a factor of two, so is the noise, as it is spread between positive and negative frequencies.
The interferogram expression is then:
\begin{equation}
I(x) = \real\left( \int_{-\infty}^{+\infty} i \S_e  e^{-2i\pi \s x + i\Phir} d\s \right).
\label{eq:interfero2}
\end{equation}

For the ideal case where $\Phir=0$, this expression shows that the interferogram is actually the real part of the Fourier transform of the spectrum at the detector, $i\S_e$. 

\subsection{Measured interferograms}

In practice, the interferogram data are discrete measurements of a real-valued signal on the detectors. Like for most FTS instruments, each data point on the interferogram corresponds to an integration of the detector while the delay line is continually in motion. This decreases the amplitude of the interferogram due to the local smearing of the fringes, but it can be kept to low values by increasing the fringe sampling.
At each OPD $\xn$, the interferogram has a measured value $\Ixn = \frac{1}{\Dx}\int_{\xn-\Dx/2}^{\xn+\Dx/2}\I(x)dx$. To first order, this has the effect of multiplying the power at each wavenumber by $\sinc(\pi\s\Dx)$. For the purpose of this paper, we consider this term to be included as part of the instrumental transmission $T_{\inst}$. Note that the value of the optical delay $\xn$ is the path difference from ZPD, not the physical location of the delay line, since there could be a multiplying factor between the two due to beam folding (e.g., for BETTII, a motion of 1~mm of the delay line creates 4~mm of OPD).



A discrete Fourier transform (DFT) is used to transform a discrete interferogram of $N$ measurements into a complex discrete spectrum with $N$ points. The resolving power of the instrument, $\R = \lambda / \D \lambda$, is dependent on the physical length scanned by the delay line $L$: $\R = L \s / 2 $ for a scan with symmetric length on both sides of ZPD. For these instruments where we scan through the whole interferogram, the data should be
sampled at least at the Nyquist rate for the interferogram response frequency of $\Dx = \lambda/2$. For a sampling exactly equal to Nyquist, we have the relationship: $N = 4 \R$.

For a double-Fourier instrument, as shown in Fig.~\ref{fig:widefield}, the ZPD for different columns on the array occurs at different delay positions $x_{\col}$, related to
the projected baseline length. The simplest way to express this is in terms of the angular offset on the sky of each column, $\xi$, along the
direction of the baseline, $\baseline$:
\begin{equation}
x_{\col} = | \baseline | \sin\xi \approx | \baseline | \xi = 48.7 \um \left({| \baseline | \over 10~\textrm{m}}\right) \ \left( {\xi \over 1~\textrm{arcsec}}\right) ,
\label{eq:delay}
\end{equation}
where we have filled in
practical units for an infrared instrument. For a far-IR interferometer working at $50~\um$, with
1-2~m diameter collectors, the delay shift across the collector point spread function (collector angular resolution) is several to ten wavelengths.
Hence the scan length to cover a wide-field array detector is comparable to the scan length required to achieve
$\R$'s of 100's to 1000's. This property is an important consideration for observation and data analysis strategies.

The ideal interferogram for a point source from a perfect instrument is an odd function of the OPD $x$, so its DFT is purely imaginary. The noise in the interferogram will be converted into spectral noise in both the real and imaginary axes so the real axis is a proportional measure of the noise. 
Referring back to Eq.~\ref{eq:interfero2}, phase shifts caused by the instrumental transfer function and source spatial visibility will
break the anti-symmetry; in practice, the DFT of a measured interferogram is complex and the real and imaginary parts are of interest.
The scientifically interesting quantities are the source spectrum and source spatial visibility: $\B$ and $\Vb$; the fixed
instrumental terms have to be calibrated or properly modeled
by observing a bright point source of known spectrum. The techniques for calibrating FTS systems are well developed
\citep[e.g.][]{Davis:2001tr}, and there are many methods proposed to correct some phase and amplitude errors \citep[e.g.][]{Forman:1966wx, Sromovsky:2003in}. 

\begin{figure}[ht!]
\begin{center}
\plotone{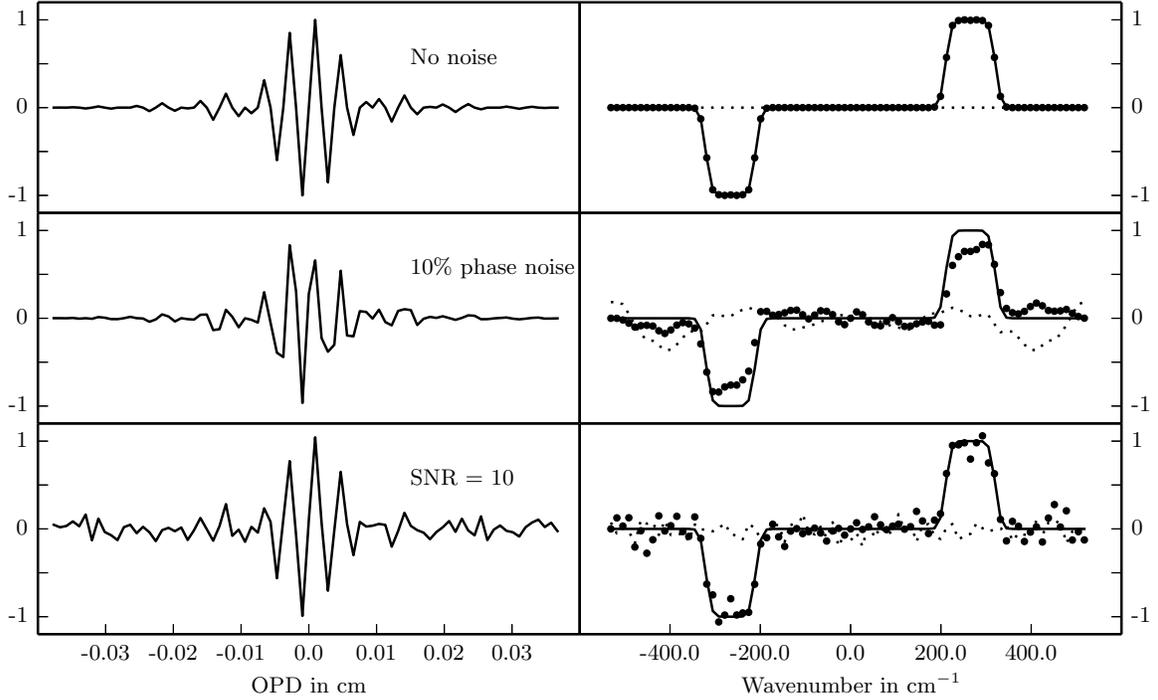}
\caption[interfero]{Effects of phase and intensity noise on the recovered spectrum (single realization of the noise). Left column: normalized interferograms, intensity as function of OPD. Right column: normalized DFT of interferograms. Solid: input spectrum multiplied by anti-symmetric transmission function; Solid circles: Imaginary part of DFT from interferogram; Dotted: Real part of DFT. First row: ideal measured signal, no noise; used for normalization of all other plots. Second row: results with a realization of phase noise of 10\% at each point of the interferogram. Third row: results with a realization of intensity noise and $\SNR_\mI=10$.}
\label{fig:interfero}
\end{center}
\end{figure}

The phase noise term $\Phir(x,\s)$ in Eq.~\ref{eq:interfero2}, and the $\SNR$ in the measured interferogram can have significant
impact on the ability to recover the source spectrum with a real instrument. The upper panel in Fig.~\ref{fig:interfero} shows an example of an interferogram (left), and the transformed $\S_e(\s_k)$ (right) for a source with flat power spectrum, multiplied by a flat bandpass function with smoothed edges.
 The middle panel of Fig.~\ref{fig:interfero} shows
the same source and instrument parameters as the upper panel, now with an assumed Gaussian OPD noise of standard deviation equal to 10\% of the central wavelength of the band $\lambda_0\equiv {1\over\s_0}$ (\textit{i.e.}, there is a $\lambda_0/10$ OPD uncertainty for each data point in the interferogram). The lower panel is the top panel observed with a incoherent background noise corresponding to $\SNR=10$ at the peak of the interferogram, and no OPD noise. 
The next sections of this paper will analyze these noise contributions and quantify their impact on the derived spectrum.

\section{Noise sources}
The two primary types of noise in a double-Fourier instrument are intensity and OPD noise. The intensity
noise consists of the astronomical and thermal background noise, the photon noise from the source, and the detector noise. The OPD noise arises primarily from uncertainties and changes in OPD, which would prevent us from accurately knowing the $x$-values of measurements in the interferogram before the FT. For convenience, we usually refer to the OPD noise as a percentage of the carrier wavelength. In the rest of this paper, a ``10\% OPD noise" signifies that the OPD for each measurement in the interferogram is known to within an error of 10\% of the carrier wavelength, or 10\% of one full fringe cycle.

\subsection{Intensity noise}
\label{sec:noisesource}
The measured signal has units of power and can be represented as the interferometric signal with additive noise:
\begin{equation}
\Im\pxn = \Ixn + \ni\pxn,
\label{eq:thermalnoise}
\end{equation}
with $\ni$ being the difference of the noise in the two outputs of the interferometer, $\ni = \nA-\nB$. When the beam combiner, optical train, and detectors are symmetric, the residual $\ni$ has zero mean. 
The total noise in $\Im\pxn$, expressed in Noise Equivalent Power, $\NEPtot$, is the sum of the three noise variances: 
\begin{equation}
\NEPtot^2 = 2\NEPph^2 +2\NEPdet^2 +  2\NEPsou^2,
\end{equation}
where $\NEPph$ and $\NEPsou$ are the noise from the background (both astronomical and thermal)
and source photon noise, respectively, in one output, and $\NEPdet$ is the noise-equivalent power characterizing each detector's noise (including phonon, readout and Johnson noise). The factor of 2 multiplies each term since we are considering the difference of both outputs.
The relation between $\NEPtot$ and the variance $\varI$ of the noise $\ni$ during an interval $\Dt$ is \citep{Sromovsky:2003in}:
\begin{equation}
\label{eq:sigI}
\varI = \frac{\NEPtot^2}{2\Dt}.
\end{equation}
For space instruments, the noise will likely be dominated by the sky background (zodiacal light, galactic cirrus emission, or optics thermal emission) and detector for a very large fraction of astronomical targets, which tend to be faint; for balloon instruments, emission from
warm optics and the atmosphere sets the noise level in the far-IR. The detector is chosen to be a small fraction of the other noise factors in order to optimize its dynamic range. 


\subsection{OPD noise}
\label{subsec:phnoise}
Observing from the ground at optical wavelengths with a double-Fourier interferometer is limited by the phase coherence 
between the apertures, which is related to the atmospheric coherence time, as discussed by \cite{Mariotti:1988vea}. The short coherence time forces fast scan rates, which degrades the sensitivity of the instrument due to short integration times and phase shifts between sequential scans. 
This is not a problem for flying platforms, since even at balloon altitudes the atmospheric coherence is not a significant 
issue \citep{Rizzo:2012jp}. The major concerns for balloon and space missions are overall instrumental stability, knowledge of ZPD, and pointing errors, which can all contribute to OPD noise.

OPD noise arises in an interferogram when the OPD at the time of a measurement is uncertain, hence compromising the reconstruction of the true $x$-value. Since this uncertainty is a physical delay $\delta_x$, 
the error in phase is wavenumber dependent: $2\pi\delta_x\s$. $\delta_x$ is the difference between the estimated $x$
and the true $x$.
For single-beam FTS instruments, internal laser metrology can provide optical path length 
measurements to high accuracy \citep[e.g.][]{Griffiths:2007uu}, and the separate paths the split beams need to travel can be kept small. 
For double-Fourier instruments, the entire optical paths upstream of the beam combiner affect the OPD, hence it is more challenging to accurately measure and estimate the OPD contributors. 
In addition, common-mode pointing errors of the collectors are geometrically converted to OPD errors. 
Hence, it is critical to know the position and orientation of the baseline vector with respect to the 
astronomical target with high accuracy in order to properly reconstruct the interferogram.

For this analysis, we identify three timescales that can be used to examine the effects of OPD noise on the interferogram. These timescales are important to consider in the design of the OPD control system of any double-Fourier interferometer. Timescale~1 is the shortest and corresponds to the integration time for a single data point, typically a few milliseconds. In practice, this kind of OPD noise could be created by high-frequency mechanical jitters in the instrument (including the delay line bearing and motor, stiction behaviors and resonant modes, reaction wheels and other self-induced vibrations...). Timescale~2 is the time it takes to acquire one single interferogram over the full field of view and at the desired resolving power, typically on the order of seconds. The sources of noise that can affect this timescale include for example pointing errors and drifts, as well errors in the knowledge of the delay line position relative to
a reference ZPD. Finally, the longest timescale to be considered, timescale~3, is the time it takes to complete one full ``track" by co-adding several consecutive interferograms to achieve the desired $\SNR$, typically a few minutes long. During this timescale, it is expected that the change in baseline orientation on the sky does not produce any significant change in the source spatial visibility function. The latter timescale is most importantly influenced by thermal variations and time-varying gradients that could change the optical alignment and mechanical configuration between the two arms.

\section{Spectral signal-to-noise ratio}
\label{sec:spectralSNR}

\subsection{Effects of Gaussian intensity noise}

In the presence of Gaussian intensity noise (thermal background and detector noise), the measured interferogram is of the form of Eq. \ref{eq:thermalnoise}. We suppose that the noise has a variance $\varI$ and zero mean, and is independent of delay position.  In particular, this assumes that the source photon noise is negligible.
The noise in the spectral domain is the transform of the noise in the interferogram domain:
\begin{equation}
\Dx\DFT(\ni) =  \Dx\sum_{n=-N/2}^{N/2-1}\ni(\xn)e^{2i\pi n k/N},
\end{equation}
where the $\Dx$ factor is to normalize the noise to a sampling bin \citep{Press:1992vya}. $n$ indexes the $N$ discrete measurements in the interferogram, and $k$ indexes the $N$ discrete wavenumbers in the spectral domain. We keep this notation throughout the paper.
The interferogram interval is symmetric with about ZPD (n=0). The noise variance is equal in the imaginary and the real domain, and can be expressed as the variance of the noise transform:
\begin{equation}
\varspec = \Dx^2\VAR\left(\real(\DFT(\ni))\right),
\end{equation}
where $\VAR$ is the variance operation. By writing out the variance we obtain:
\begin{equation}
\varspec = \Dx^2\varI\sum_{n=-N/2}^{N/2-1}\cos^2(2\pi nk/N) =\frac{N}{2}\Dx^2\varI ,
\end{equation}
where we used $\sum_{n=-N/2}^{N/2-1}\cos^2(2\pi nk/N) = N/2$ for $k~\neq~0$. 

The signal at wavenumuber $\s_k$ in the discrete spectrum $\S_e(\s_k)$ observed at the detectors is:
\begin{equation}
 \S_e(\s_k) = \frac{1}{\delta\s}\int^{\s_k+\delta\s/2}_{\s_k-\delta\s/2} \S_e(\s)d\s,
\label{eq:signal}
 \end{equation}
where $\delta\s = (N\Dx)^{-1}$. A spectral line of power $P_e$ at $\s_{k_0}$  will thus have an apparent flux density $\S_e(\s_k) = N\D xP_e$ at $k=k_0$ and $0$ for all other $k$. The signal-to-noise ratio in the spectrum can be expressed in general as:
\begin{equation} 
\SNR_k  = \frac{\S_e(\s_k)}{\sigspec} =\sqrt{\frac{2}{N}}\frac{\S_e(\s_k)}{\Dx\sigI} .
\label{eq:spectralSNR}
\end{equation}



Defining the central wavenumber of the band as $\s_0$, the spectral resolving power of the transformed interferogram is $\R = \Dx N\s_0/2$. We introduce the sampling parameter $\samp = (\s_0\Dx)^{-1}$ which is the number of data samples per fringe for the central wavenumber in the band. The spectral resolving power at the band center can now be written $\R = \frac{N}{2\samp}$.  In practice one wants to pick a value of $\samp$ that ensures Nyquist sampling on the fringe for all wavenumbers in the band so $\samp \sim 3$ or greater is typically preferred. For a given integration time per data point (given $\SNR_\mI$), increasing the fringe sampling effectively increases the amount of time spent on the fringe, so the spectral $\SNR$ should increase with $\sqrt{\samp}$. Note that as long as we Nyquist-sample the fringe, there is no difference between multiplying the fringe sampling by some factor, and increasing the integration time per data point by the same factor, since in both cases the effective time on the fringe is equally increased.  

For continuum and low-spectral resolution measurements, it is useful to relate $\SNR_k$ to the $\SNR$ in the interferogram at the location of maximum intensity of the fringe, using physical quantities. The noise in each discrete measurement of the interferogram is $\sigI$. The signal at maximum intensity is $\mI_\textrm{max} = \D\s\oS$, where $\D\s$ is the width of the bandpass filter and $\oS$ is the average value of the signal in the band. Defining $\SNR_\mI = \mI_\textrm{max}/\sigI$, and noting that $\sqrt{N\Dx^2/2} = \frac{1}{\s_0}\sqrt{R/s}$, we obtain:
\begin{equation}
\SNR_k = \frac{\S_e\sqrt{2}}{\sqrt{N}\Dx\sigI} = \frac{\S_e(\s_k)}{\oS}\sqrt{\frac{s}{\R}}\frac{ \s_0}{\D\s} \SNR_\mI .
\label{eq:SNRratio}
\end{equation}
Thus for a given $\SNR_\mI$, the $\SNR$ in a channel of the final spectrum depends inversely on the square root of the resolving power $\R$ and the fractional bandwidth $\frac{\D\s}{\s_0}$; and it depends directly on the square root of the number of samples per fringe $\sqrt{\samp}$. 

For a given integration time, $\SNR_\mI$ is proportional to $\sqrt{\frac{\D\s}{\s_0}}$, which means that $\SNR_k$ is inversely proportional to $\sqrt{\frac{\D\s}{\s_0}}$. Hence, maximizing the $\SNR$ in the interferogram by increasing the bandwidth does not lead to a better spectral $\SNR$. Although the central fringe has more $\SNR$ with a larger bandwidth, the fringe envelope is decreasing more rapidly, and we see less fringes with good $\SNR$. With a smaller bandwidth, the envelope is broader and although each fringe has less $\SNR$, we are seeing more of them within the scan, so we spend more time in an interesting region of the interferogram. The number of fringes within the main lobe of a standard sinc function is indeed inversely proportional to the fractional bandwidth. This is an important concept that can be used to tune the spectral resolving power appropriately: if the instrument has a large bandwidth, increasing the scan length (hence $\R$) well beyond the fringe pattern's main lobe will hurt the sensitivity, since many data points will be adding noise and almost no signal. The dependencies discussed above can be summarized as:
\begin{equation}
\SNR_k^2 \propto \left(\frac{s}{\R}\right)\left(\frac{ \s_0}{\D\s}\right)  \Dt .
\end{equation}


\subsection{Effects of Gaussian OPD noise}

This section derives analytic expressions for the effects of Gaussian-distributed OPD noise. We look at the general case in order to derive sensitivities for double-Fourier instruments. Here, we suppose that the OPD from the delay line, the OPD within each arm of the instrument, and the OPD caused by an off-axis source are all measured or estimated with some residual error. Hence, the data points measured in the interferogram are associated with a delay value relative to ZPD, and if necessary, resampled to produce an evenly-spaced delay axis. This is necessary to use the FT and retrieve the spectrum. The noise on the delay estimate can be
characterized as a wavenumber-dependent phase error in the interference on the two beams. In the following, we quantify the impact of this noise on the spectral $\SNR$, in order to understand how good our knowledge of the OPD needs to be to make sure the OPD noise effects are not dominant.



Let's consider a single frequency signal first, so that the phase is proportional to the OPD. 
If we suppose that these residual phase errors $\Phir(x)$ are represented by a Gaussian distribution with zero mean and variance $\varPhir$, then the primary effect of the noise is to change the instantaneous power in $\I(x)$ by the factor $e^{i\Phir(x)}$. Now we consider a large ensemble of realizations of this noise distribution in order to predict its effect on the $\SNR$. Using the expression from \cite{Richards:2003bp}, for sufficiently small phase errors ($<\pi$ radians), the intensity of the coherent signal is reduced, on average, by a factor $e^{-\varPhir/2}$. For Gaussian-distributed OPD uncertainties with standard deviation $\lambda/20$, where $\lambda$ is the wavelength, the signal intensity is reduced by 5\%; for $\lambda/10$ the amplitude is reduced by 18\%. To give a practical example of the impact of this effect, we can consider the case of BETTII: if we assume that the uncertainty in the attitude of the payload is the only source of OPD noise, then knowing the attitude to within 0.1" rms will reduce the signal, on average, by 18\% at 40~$\um$.

For the polychromatic case, the delay position uncertainty, $\delta_x$, creates larger phase errors the shorter the wavelength, 
$\Phir(k) = 2\pi\delta_x \s_k$. A given OPD error distribution of variance $\varopd$ yields a degradation across the band, $e^{-\varPhir(k)/2}$, with $\varPhir(k) = (2\pi)^2\varopd\s_k^2$.

Of course, the power lost from the coherent fringe pattern is still present in the scan; 
it becomes part of the incoherent signal seen by each output. 
In the limit where there is no spectral noise from the background or detectors, defining $\S_k\equiv\S_e(\s_k) $ we have:
\begin{equation}
\SNR_k= \frac{\S_k e^{-\varPhir(k)/2}}{\sqrt{\frac{1}{2\samp\R}\sum_{k'} \left[\S_{k'}^2(1-e^{-\varPhir(k')})\right]}},
\label{eq:noiseph}
\end{equation}
where $k'$ designates an index on all positive wavenumber bins. Note that $N=2\samp\R$. This relationship is identical to the one derived by \citet{Meynart:1992fv}, and we suggest an alternate and more detailed justification for it (see Appendix \ref{ap:phasenoise}). Studying this relationship, all the wavenumbers contribute to the white noise at a given wavenumber $\s_k$. The strongest lines (strongest $\S^2_{k'}$) and the shortest wavelengths (strongest $1-e^{-\varPhir(k')}$) contribute the most to the overall noise.
To summarize, considering an ensemble average of interferograms, OPD noise degrades the spectral $\SNR$ in two ways: first, it reduces the overall signal in the interferogram; second, it converts this lost power into white noise.


More realistically, observations will have
both intensity and OPD-generated spectral noise. In this case, the intensity noise and the scattered power
add in quadrature to give:
\begin{equation}
\SNR_k = \frac{\S_k e^{-\varPhir(k)/2}}{\sqrt{\frac{1}{2\samp\R}\sum_{k'} \left[\S_{k'}^2(1-e^{-\varPhir(k')})\right] + \samp\R\Dx^2\varI}}.
\label{eq:noisephth}
\end{equation}

The numerator of Eq.~\ref{eq:noisephth} shows that any amount of OPD noise will reduce the spectral $\SNR$. However, the impact of OPD noise is even greater when the power lost from the fringe is comparable to the intensity noise, as the first term of the denominator starts to matter. In fact, for arbitrarily large source fluxes, this equation reaches an asymptotic value which depends only on the OPD noise, and sets the maximum $\SNR$ achievable on average in a single scan. This is relevant for astronomical calibrators which can be so bright that the intensity noise term is negligible. In that case, assuming constant OPD noise, more $\SNR$ is only achievable by co-adding consecutive scans, as we discuss in the next section and in Appendix C. For most astronomical applications, where targets are usually faint compared to the intensity noise, it is expected that the first term of the denominator will be negligible.

\subsection{Co-adding interferograms}

Eq.~\ref{eq:noisephth} is the general case of a single interferogram with OPD and intensity noise. In practice, we would co-add $M$ interferograms in one ``track" to build up $\SNR$, but this puts stringent requirements on the performance of the control system and OPD estimator, because consecutive interferograms need to stay aligned with each other to within a small fraction of the carrier wavelength, to avoid causing OPD noise. The design and performance of the OPD estimator is highly implementation-specific, but most balloon and space designs will likely include an estimator that either directly measures the OPD, or indirectly infers it from the measurement of another quantity. 

A direct OPD measurement can be achieved for example with a fringe-tracking instrument, while an indirect OPD estimate can be an attitude measurement, which can be related to the OPD by simple geometry by using some assumptions. The latter scheme only works if the OPD errors are only influenced by pointing uncertainties over the timescale of a track, and that all other OPD contributors are modeled and corrected with comparatively high fidelity. The spectral $\SNR$ over $M$ scans can be determined from Eq.~\ref{eq:noisephth} by multiplying the whole equation by a factor of $\sqrt{M}$. The OPD noise term causing the phase noise variance $\varPhir$ then corresponds to the variance of the OPD uncertainties for each point of a scan, plus the variance of the OPD estimation error in determining the position of the center of each scan, which is necessary to properly co-align them (Appendix C).

\subsection{Implications for spectroscopy}
A primary application for BETTII and proposed missions like SPIRIT will be the measurement of the spectral energy distribution
from warm dust associated with star formation in different environments. These types of measurements require broad wavelength
coverage but not especially high spectral resolution since the emission can be characterized as a sum of Planck functions over
a range of temperatures. For an instrument like BETTII, covering from 30-55~$\um$ and 60-90~$\um$ simultaneously,
$\R\sim 10$ in each band is sufficient to accomplish much of the science.

Spectral measurement with $\R\sim 10$ requires covering a delay range of $\pm 10~\lambda_0$ for a single source. On the other
hand, a delay range of 35-70~$\lambda_0$ (see Eq.~\ref{eq:delay}) is needed to move ZPD across 1 arc-minute of sky. Hence, typically,
the delay requirements for spatial coverage creates interferograms with higher resolution than needed to measure the continuum, and the full scan needs to be cut into smaller arrays around each target in the field. The size of these smaller arrays depends on the desired spectral resolving power $\R$, and the required sensitivity, as shown in Eq.~\ref{eq:SNRratio}. However, the additional data can be used for higher-resolution spectroscopy, for example to measure specific atomic lines in the far-IR. The $\SNR$ for lines is actually increasing with the square root of the number of data points in the interferogram, as the broadband noise gets more diluted in increasingly narrower spectral bins (see Eq. \ref{eq:signal}, \ref{eq:spectralSNR}). 


As discussed for FTS instruments \citep[e.g.][]{Davis:2001tr},
apodization, the weighting of the points of the measured interferogram before applying the DFT, is one method for optimizing the $\SNR$
in the spectrum.
 The weight scheme is optimized to measure a specific type of spectrum: narrow line, broad features, continuum. 
The method relies on the fact that the data points close to the center or edges of a fringe packet contain information about low or high spectral frequencies, respectively. For example, if the purpose of an observation is to study continuum, it is appropriate to apply smaller weights to data points far away from the central fringe, since they add noise and very little $\SNR$. 

A common low-resolution spectroscopy case can be derived analytically
if a source has a spectrum following a power law distribution over the covered band. We can 
write $\S(\s) \propto \s^\alpha$ where the exponent $\alpha$ is the quantity of interest. 
Several methods have been developed to properly fit these power laws using maximum entropy and other 
techniques \citep[e.g.][]{Clauset:2007iy}. Here we use a simple estimator and provide 
a ready-to-use formula to help quantify the sensitivity of double-Fourier instruments.

By taking the logarithm of the spectrum, the problem is turned into a weighted linear fit in log-log space, where we want to determine the slope of a line. The noise in the new domain is $\sigL = \left|\frac{d(\ln(\S))}{d\S}\right|\sigspec = \sigspec/\S = 1/\SNR_\S$. The weights $\wk = 1/\sig_k^2$ of the linear fit are then simply the values of the spectral $\SNR$ squared at each data point, $\SNR^2_k$. The error on the weighted least square estimate of the slope is \citep{Bevington:2003tc}:
\begin{equation}
\sigalpha^2 = \frac{\sum \wk}{\sum \wk\sum \wk X_k^2 - \left(\sum\wk X_k\right)^2},
\end{equation}
where $X_ k \equiv \ln(\s_k)$ is the natural logarithm of the wavenumber for data point $k$. In the case of uniform spectral signal-to-noise ratio $\SNR_\S$ over $m$ points of the spectrum, this expression simplifies to:
\begin{equation}
\sigalpha^2 = \frac{1}{m\times\SNR^2_\S\times\VAR(X_k)}.
\end{equation}
This equation indicates that the variance of the spectral index estimate decreases with the number of points used to calculate the estimate, the spectral $\SNR$ squared, and the variance of the points distribution on the logarithmic wavenumber axis. For example, for 10 spectral points spread evenly from 30 to 55~$\um$, each with a spectral $\SNR$ of 5, we obtain an error on the slope determination $\sigalpha\sim 0.3$.

\section{Spectral sensitivity analysis for BETTII}
\label{sec:implications}
This section applies elements of the above discussion to BETTII. A general discussion on the details of BETTII can be found in \cite{2014PASP..126..660R}.
On BETTII, two mirrors collect light with an altitude-azimuth pointing system. The truss that holds the two mirrors moves in azimuth and determines the baseline vector, while the mirrors themselves move only in elevation. While BETTII does not physically rotate about the line of sight to cover different baseline angles, the payload always stays horizontal and the projection of its baseline vector changes as a source moves across the sky, hence covering different angles in the ($u, v$)-plane. The absolute OPD and ZPD of the instrument cannot be
measured, maintained, or known with perfect accuracy, especially during the flight itself, due to attitude estimation errors leading to our inability to perfectly estimate the orientation of the baseline vector in real time. In fact, a significant component of the mission's
design and implementation involves the selection and coordination of the suite of instruments which provide attitude measurements to construct the OPD estimator.

A second relevant aspect of BETTII is that the detectors are cryogenic bolometers \cite[see][for similar architectures]{2014ApJ...790...77S} with 1/f noise which
sets an optimal read-out time for the detectors of around 2.5 milliseconds (timescale 1). With BETTII's designed field coverage
of 2 arcminutes, full field scans consist of 1024 points and take 3 seconds to complete (timescale 2). Due to thermal emission from the atmosphere,
warm mirrors, and cryostat windows, BETTII will be in the background noise limited case for all science targets.
It is anticipated that 200 scans will typically be co-added to create one single visibility measurement over 10 minutes (timescale 3). 
For most source locations, the variation of the baseline orientation due to change in parallactic angle is not significant over this period.

\subsection{Noise sources and control system}
Table~\ref{tab:noise} shows our estimates of the
background power levels associated with the atmosphere, warm optics, and windows in the two BETTII bands. 
The detectors themselves have been designed to have a noise level comparable to the background to optimize the use of the
dynamic range of the devices. The total NEPs of the short and long bands are expected to be $\sim 2\times 10^{-15}$ W.Hz$^{-0.5}$ and $\sim 1\times 10^{-15}$ W.Hz$^{-0.5}$, respectively. The source photon noise is negligible compared to the total NEP. 

Balloon instruments are subject to low frequency ($<~0.5$~Hz) pendulum modes and other oscillations introduced by the system's geometry and mass distribution, which make pointing a challenge. However, it is expected that the balloon environment is free of perturbations at any higher frequency (other than the instrument specific perturbations). Hence, sensors with high electrical bandwidth can robustly estimate the pendulum modes to gain accurate knowledge of the attitude, which can be used as our indirect OPD estimator since it is geometrically related to the phase on sufficiently short timescales.

The BETTII control system is organized with three different levels of control loops \citep{2014SPIE.9143E..3HR}: the coarse pointing loop, the fine pointing loop, and the OPD loop. The coarse pointing loop uses gyroscopes and star cameras to keep the baseline oriented within 10-15" of an appropriate near-IR guide star. A dichroic splits the near-IR (1-2$~\um$) from the far-IR (30-90$~\um$) inside the cryostat before the scanning delay line. The guide star is imaged through each of the two arms on two separate readout windows of a near-IR detector array that shares most of the optical path with the science channels. The fine control loop uses fast-steering tip-tilt mirrors, located at the pupils of each arm, to control the guide star image on each window and maintain good overlap of the beams at the science detectors. This loop reads the near-IR detector and generates a tip/tilt correction at 100~Hz. We expect to achieve beam overlap to within better than 1.5" at all times when a guide star is available. The spatial resolution of an individual BETTII beam
is 17" in the short wavelength band so this is a little better than 1/10th of a resolution element. The interferometric visibility loss
for this overlap error is anticipated to be less than 0.5\%.

We do not expect to be able to maintain the three dimensional orientation of the truss, and hence the baseline
vector, to much better than 10" rms, due to the various pendulum modes mentioned above and large inertia of the payload.
However, the errors in OPD introduced by pointing errors can be corrected directly using a delay line. BETTII uses a delay line external to the cryostat to correct the OPD at the entrance of the cryogenic volume. This delay line is completely separate from the science delay line which scans the OPD to produce the interferogram. Two delay lines are not a requirement for a double-Fourier instrument in general as the job can be done in theory by a single mechanism, with sufficient range and mechanical bandwidth. The external delay line on BETTII allows for the possible future upgrade
of correcting and monitoring the OPD outside of the cryostat using the near-IR channel by implementing a fringe tracker \citep{Rizzo:2012jp}.

For the OPD loop on BETTII, the angles of the tip/tilt mirrors which are used to maintain overlap of the beams act as an estimator of the baseline orientation, and hence as an indirect estimator of the OPD. The attitude estimates computed from these angles are converted to OPD and fed to the external delay line so that the OPD at the entrance of the cryostat stays as constant as possible. Because the pendulation modes have periods of a few to tens
of seconds and should be well-behaved, we expect to be able to trust the control signals and estimate the attitude of the baseline vector to $\sim 0.12$" rms, which corresponds to a fifth of a detector pixel in the near-IR tracking array. A 0.12" attitude error indirectly corresponds to a delay uncertainty of 5~$\um$, or 12\% of a wavelength at 40~$\um$. This is a critical consideration when co-adding consecutive interferograms. With this amount of OPD noise we expect, on average, a $\sim 25\%$ degradation in $\SNR$ for all sources in the short band, simply from the effects of phase noise in reducing the coherent signal (see Eq. \ref{eq:noisephth}).

Even with a stable OPD estimator, the absolute ZPD of the instrument must be measured during flight and tracked over long timescales as the instrument and the truss cool down to ambient temperatures ($\sim$240~K). This can be accomplished by observing a bright point source with known position periodically during a flight and identifying the center of the interferogram response (see Appendix C).

\subsection{Derived sensitivity and faintest detectable targets}

Incorporating these sources of noise with the formulas derived in the previous sections leads to the sensitivity values shown in Table~\ref{tab:sensitivity}. In this table we show the sensitivity in the two bands. The minimum detectable flux density (MDFD), which is the flux that provides $\SNR_\mI = 1$ in a single interferogram, is 15~Jy and 26~Jy in band 1 and 2 respectively. For 200 scans averaged with a OPD noise between scans of $5~\um$, the MDFD is 1~Jy and 2~Jy, using a matched filter efficiency of 0.5 and 0.4, respectively \citep{Mighell:2005fwa}. The faintest detectable spectroscopic point source that leads to a spectral $\SNR=5$ is 25~Jy and 13~Jy, respectively. These are determined for ``normal observing", which consists of co-adding 200 scans in 10 minutes that span the whole 2'x2' field of view, using a spectral resolution of $\R=10$ and a nominal OPD noise of $5~\um$~rms. 

At the bottom of the table, we also show the results in case we were using the instrument in an ``enhanced sensitivity" mode. This mode is mentioned here to illustrate the flexibility of the interferometer and its observing modes. It consists of increasing the individual integration time for each point in the interferogram by a factor of 3, while reducing the interferometric field of view by the same factor of 3: while the intrinsic field of view is unchanged at the detector, for the same scan time we only cover enough OPD range to cross ZPD for a subset of the pixels of the detector (and obtain a scan of the same length). This mode could be used for example for isolated targets which are located in less crowded star fields, by optimizing the time spent close to ZPD, where there is more signal (as we are interested in low-resolution spectroscopy). BETTII's observing parameters can be changed during flight so that the instrument stays flexible to optimize the chance of seeing fringes.

Finally, we show the overall sensitivity as a function of point source flux density (Eq.~\ref{eq:noisephth}) for both observing modes and both bands in Figure~\ref{fig:SpectralSNR}. In normal background-limited regime, the sensitivity curves should be straight lines. Here, OPD noise creates a decrease in overall sensitivity as a reduction in coherent power, but also, for brighter targets, from the power lost from the fringe that is converted to white noise (which causes a deviation from straight lines). For very bright targets of 50~Jy or more, it is possible to measure the OPD accurately within each interferogram by tracking the fringes in the science channels themselves (see Appendix C). For sufficiently large $\SNR$, this process has less error than the assumed $5~\um$ OPD noise coming from the indirect OPD estimation, so the OPD noise decreases for these very large fluxes to become negligible. This is particularly attractive for in-flight testing and calibration.

It is important to note that for sufficiently faint targets, it is impossible to accurately measure the OPD using single scans or co-adds of scans: we rely on the OPD estimator to have sufficient stability to properly co-add scans until the next calibration measurement. This needs to be considered carefully when planning the observation strategy, as long stretches without calibration could lead to a total loss of the OPD information (hence a total loss in scientific data), due to other OPD noise contributors such as thermal drifts that impact the payload on long timescales.
\section{Conclusion}


Spatio-spectral interferometry can enable high resolution spectral imaging of wide fields at
far-IR wavelengths. Implementation of the technique provides some new instrumental
challenges compared to traditional Fourier Transform Spectroscopy, such as the fact that the measured spectrum is a mix of the source's spectral and spatial information.


In a double-Fourier system, the zero path difference for each detector pixel occurs at a different delay setting of the delay line. The delay stroke needed to cover a scientifically interesting field of view is equivalent to a spectral resolving power of 100's to 1000's for the central pixels.

We present an analysis of the impact of Gaussian intensity and OPD noise on the spectral sensitivity.
Intensity noise, essentially thermal noise from the optics, sky, astrophysical background, and detector, is similar to noise in 
FTS systems with the exception that the longer scan lengths required to cover the spatial
field add noise; this can be mitigated by cutting the interferogram for each pixel  into smaller arrays centered on each source's ZPD to match the desired spectral resolving power, and by apodizing the interferogram to increase sensitivity to the spectral properties of interest. We show that the spectral $\SNR$ is inversely proportional to the square root of the spectral resolving power and the square root of the fractional bandwidth, in contrast to the $\SNR$ on the central interferometric fringe which is directly proportional to the latter.

OPD noise is not usually relevant for FTS systems, but is intrinsic to double-Fourier instruments, since the two incoming beams go through long separate paths before combination. For instruments on balloons or in space, the OPD noise is expected to be dominated by disturbances from the instrument and from pointing errors. On average, OPD noise reduces the coherent power in the interferogram, and converts the power lost from the fringe into additional white noise in the spectrum. We argue that there are three relevant noise timescales: the time to take a single data point, the time to collect a complete interferogram, and the time to co-add $M$ interferograms together in a track. The latter corresponds to the timescale that the source spatial visibility function changes significantly, due to the rotation of the baseline angle on the sky.

We derive the spectral sensitivity of double-Fourier instruments to intensity and OPD noise. The expressions in this paper are derived in the general case and can be used to design any instrument that implements this method.

Applied to the case of BETTII, these equations lead to spectral sensitivity estimates of 25 and 13~Jy in its 30-55~$\um$ and 60-90~$\um$ bands, respectively, to achieve a spectral $\SNR=5$ in 10 minutes with $\R=10$ and an assumed OPD noise of $5~\um$ rms.

\acknowledgments

The material presented in this paper is based upon work supported by NASA Science Mission Directorate through the ROSES/APRA program, with additional support from NASA Goddard Space Flight Center, and NASA GSFC grant NNX11AG92A to the University of Maryland. Work by T. Veach was supported by an appointment to the NASA Postdoctoral Program at GSFC, administered by the Oak Ridge Associated Universities under contract with NASA. We would like to thank the anonymous referee for suggested improvements to the paper.

\appendix
\section{Deriving the Interferogram Equation in a Double Fourier System}
\label{ap:interfero}
The interferogram from a double-Fourier system is different from the interferogram
for an FTS in several ways that derive from the fact
that the double-Fourier system starts with two independent input beams viewing the same astronomical target. For this derivation,
we will follow the convention in the FTS literature and consider the propagation of a single plane wave (radiation from a point source at infinity) at wavenumber $\s\equiv{1/\lambda}$ through the system.

Figure 2 in the main text shows the setup for a typical double-Fourier system with the K-mirror on one
arm to keep the sky images at the same rotation on the two paths, and the delay line
in the other arm to allow adjustment of the relative path lengths between path 1 and 2.
The plane wave travels a distance $x_1$ on path 1 from an entrance aperture an arbitrary distance
above the siderostat to the beam combiner: $a_1(\s) e^{-2\pi i \s x_1+\phi}$,
where $a_1$ is the amplitude of the electric field and $\phi$ corresponds to an arbitrary phase offset. For convenience of notation, in the following derivation we drop the amplitudes' dependence on wavenumber by writing $a_1$ instead of $a_1(\s)$.

The wave also undergoes phase shifts caused by reflections and partial reflections along
the path. A full reflection for light traveling in air or a vacuum causes a 180~$\deg$ phase shift;
a 50\% reflection at the beam splitter/combiner causes a 90~$\deg$ phase shift between reflected and transmitted beam \citep{2000plbs.conf.....L}. Since the instrument
measures the combined light at the detectors, what matters is the difference in the
numbers of reflections along path 1 and 2. In the case of the particular BETTII implementation, path 1 contains one more reflection than path 2.

The electrical fields arriving at the ``+" and ``-" detectors are then:
\begin{eqnarray}
A_- &=& a_1 e^{-2\pi i \s x_1+ i \pi + i\pi /2+\phi} + a_2 e^{-2\pi i \s x_2 +\phi },\\
A_+ &=& a_1 e^{-2\pi i \s x_1 + i \pi +\phi} + a_2 e^{-2\pi i \s x_2 + i\pi /2 +\phi},
\end{eqnarray}
where the $\pi$ phase shift on path 1 occurs because there is one extra reflection compared to path 2 (see Fig.~\ref{fig:optics}), and $\phi$ corresponds to an arbitrary phase offset. The detectors are power detectors so defining the intensity $ I = A^* A$:
\begin{eqnarray}
I_- &=& a_1^2 + a_2^2 + a_1 a_2 \left(e^{-2\pi i \s (x_1 - x_2) + 3i\pi /2} + e^{2\pi i \s (x_1 - x_2) - 3i \pi/ 2}\right),\\
I_+ &=& a_1^2 + a_2^2 + a_1 a_2 \left(e^{-2\pi i \s (x_1 - x_2)  + i\pi /2} + e^{2\pi i \s (x_1 - x_2) -  i \pi/ 2}\right).
\end{eqnarray}
Defining $x \equiv x_1 - x_2$ and expanding the complex exponentials, the equations can be simplified to:
\begin{eqnarray}
I_- &=& (a_1^2 + a_2^2 ) \left( 1 - {2 a_1 a_2 \over a_1^2 + a_2^2} \sin( 2 \pi \s x) \right),\\
I_+ &=& (a_1^2 + a_2^2 )\left ( 1 + {2 a_1 a_2 \over a_1^2 + a_2^2} \sin( 2 \pi \s x) \right),
\end{eqnarray}
where $x$ is now the difference in the physical length between the two light paths.
For the case of equal wave amplitudes on path 1 and 2 ($a_1=a_2=a$):
\begin{equation}
I_\pm = 2 a^2 ( 1 \pm \sin( 2 \pi \s x) ).
\end{equation}
The generalization of this equation to a source distribution on the sky requires the recognition that $a_1$ and $a_2$
are complex values such that $|a_1|^2(\s)$ and $|a_2|^2(\s)$ are
power from the source at wavenumber $\s$, while $a_1 a_2^*$ is the correlated power seen through the two apertures which is
the source spatial visibility, $\gamma(\baseline,\s)$, and is in general a complex valued function. $\gamma(\baseline,\s)$, which is a function of
the baseline vector $\baseline$ connecting the two light collectors, and $\s$, is the Fourier transform of the
source emission distribution on the sky.
For the general case, the previous equations become:
\begin{eqnarray}
I_- & = & |a_1|^2 + |a_2|^2 + \gamma(\baseline,\s) e^{-2\pi i \s (x_1 - x_2) + 3i\pi /2} + \gamma^*(\baseline,\s) e^{2\pi i \s (x_1 - x_2) - i 3\pi/ 2},\\
I_+ & = & |a_1|^2 + |a_2|^2 + \gamma(\baseline,\s) e^{-2\pi i \s (x_1 - x_2) + i\pi /2} +  \gamma^*(\baseline,\s) e^{2\pi i \s (x_1 - x_2) -i  \pi/ 2}.
\end{eqnarray}
The same simplification as before can be done except that $\gamma(\baseline,\s)$ is a complex-valued function. If we define the normalized spatial
visibility as
\begin{equation}
\Vb(\s) = {2\gamma(\baseline,\s)\over a_1^2 + a_2^2},
\end{equation}
then the equation for $I_\pm$ becomes:
\begin{eqnarray}
I_\pm & = & ( |a_1|^2 + |a_2|^2 ) \left[ 1 \pm (\real\left(\Vb(\s)\right) \sin(2 \pi \s x) - \imag\left(\Vb(\s)\right) \cos(2 \pi \s x))\right],\\
I_\pm & = & ( |a_1|^2 + |a_2|^2 ) \left[ 1 \pm \real\left( i \Vb(\s) e^{-2 \pi i \s x}\right)\right],
\end{eqnarray}
where $\real(f)$ is the real component of $f$ and $\imag(f)$ is the imaginary component.

The same style of derivation can be done with for a realistic instrument with a complex transfer function.  If
we characterize the spectral transmission function as $t_1(\s) = |t_1(\s)| e^{i\Phi_1(\s)} $ along path 1, and
$t_2(\s) = |t_2(\s)| e^{i\Phi_2(\s)} $ on path 2, then the amplitude mismatch of the spectral transmission function in each path reduces the power in the interferogram and
the phase differences introduce a phase factor $\Phi_{i} = \Phi_1 - \Phi_2$ into the exponential term.
As a result, the source visibility in the previous equations is multiplied by a normalized, instrumental visibility loss term, $\Vi = |\Vi(\s)|e^{i\Phi_i(\s)}$:
\begin{equation}
I_\pm = ( |t_1|^2 |a_1|^2 + |t_2|^2 |a_2|^2 ) \left[ 1 \pm \real( i \Vb(\s)\V_i(\s) e^{-2 \pi i \s x})\right].
\end{equation}

\section{Spectral noise in presence of gaussian phase noise}
\label{ap:phasenoise}

Suppose that the signal is a line of power density $2\S$ centered on bin number $k$ corresponding to wavenumber $\s_k$. In the complex interferogram, the line has a power density $\S$ in bin $k$ and $-\S$ at $-k$, and zero everywhere else. To simplify the analysis, let's focus on the positive frequencies, which only contain half the noise. The interferogram at delay $\xn = n\Dx$ is $\Ikxn =\S\dsig e^{-2i\pi \s_k\xn}$. Through a simple DFT, the value of the line in the spectrum in ideal conditions is:
\begin{equation}
\Dx\DFT(\Ikxn)[k'] = \Dx\sum_{n=-N/2}^{N/2-1}\S\dsig e^{-2i\pi \s_k\xn }e^{2i\pi n k'/N} = \Dx\sum_{n=-N/2}^{N/2-1}\S\dsig e^{-2i\pi (k-k')n/N },
\end{equation}
which is equal to $\Dx N\S\dsig = \S$ for $k=k'$ and zero everywhere else.
Note that we have $\s_k\xn = k\dsig n\Dx = kn/N$. and $\dsig = (N\Dx)^{-1}$. The phase noise degrades the effective power of the line, so it is now $\S e^{-\varPhir/2}$ \citep{Richards:2003bp}. The noisy interferogram is $\Ikxn =\S\dsig e^{-2i\pi kn/N}e^{i\Phir\pxn}$.

Designating the operator $\langle \rangle $ as the ensemble average, the noise $\varspec$ in the interferogram is the variance of the DFT:
\begin{eqnarray}
\varspec [k'] & = & \VAR(\Dx\DFT(\Ikxn)[k']) \\
& = & \Dx^2\left(\left\langle\left\vert \sum_n \Ikxn e^{2i\pi n k'/N} \right\vert^2\right\rangle - \left\vert\left\langle \sum_n \Ikxn e^{2i\pi n k'/N} \right\rangle\right\vert^2\right) ,\\
& = & \Dx^2\left(\sum_n\sum_{n'}\left\langle \Ikxn\Iksxn\right\rangle e^{2i\pi (n-n') k'/N} - \sum_n\sum_{n'}\left\langle \Ikxn\right\rangle\left\langle\Iksxn\right\rangle e^{2i\pi (n-n') k'/N}\right), \\
& = & \Dx^2\sum_n\sum_{n'} \left[ \left\langle \Ikxn\Iksxn\right\rangle - \left\langle \Ikxn\right\rangle\left\langle\Iksxn\right\rangle\right] e^{2i\pi (n-n') k'/N}.
\end{eqnarray}
We can write $\langle \Ikxn\Iksxn\rangle = \langle \S^2\dsig^2 e^{-2i\pi (n-n') k/N} e^{i(\Phir\pxn - \Phir\pxnp)}\rangle$. This quantity is equal to  $\S^2\dsig^2 e^{-2i\pi (n-n') k/N} e^{-\varPhir}$ when $n\neq n'$ and equal to $\S^2\dsig^2$ when $n=n'$. The quantity $\langle \Ikxn\rangle\langle\Iksxn\rangle$ is equal to $\S^2\dsig^2 e^{-2i\pi (n-n') k/N} e^{-\varPhir}$ for all $n$ and $n'$. Hence, the term in the sum is nonzero only for $n=n'$, for which it is $\S^2\dsig^2(1 - e^{-\varPhir})$. The value of the sum is then:
\begin{eqnarray}
\varspec [k'] & = & \Dx^2\sum_n\S^2\dsig^2(1-e^{-\varPhir}), \\
& = & \Dx^2N\S^2\dsig^2(1-e^{-\varPhir}),\\
& = & \frac{1}{N}\S^2(1-e^{-\varPhir}).
\end{eqnarray}
This quantity is independent of $k'$, so the noise is white. The negative frequencies contribute the same amount, doubling the noise variance. However, we are only considering the imaginary part of the spectrum, so only half the noise variance is important in our calculation of our $\SNR$. The last expression thus represents the variance of the noise that is useful for our $\SNR$ calculations.

\section{Fringe tracking in the science channels}

For sufficiently bright sources, it is possible to self-calibrate the OPD between subsets of the $M$ interferograms in a track, to prevent the drift of an indirect OPD estimator. The idea is to bin consecutive interferograms in subsets in order to build up enough $\SNR$ to clearly see a fringe and be able to estimate its position with sufficient accuracy. Then, the different subsets within a track can be offset and co-added with better accuracy (smaller OPD noise) than if we were co-adding the $M$ interferogram individually with only the instrument OPD estimator noise. The best scenario would be when the fringe has a high $\SNR$ in each single interferogram - which will be the case of calibrators for BETTII.

There are many ways to fit the location of the fringe center, and the error associated with each method is highly implementation-specific. Here, we consider the simple example of a fringe tracking algorithm in two steps \citep{Rizzo:2012jp}: a Hilbert transform of the interferogram to obtain its envelope; and a centroid of the points of the envelope above a certain $\SNR_\mI$ threshold. The Hilbert transform doubles the error variance in the interferogram, and in the worst case, the centroid has an error variance of approximately $ (n\times\SNR_\mI^2)^{-1}$, where $n$ is the number of data points above the threshold $\SNR_\mI$. The conversion to a phase leads to a phase error variance equal to $[\varPhir(\s)]_\textrm{direct} \sim 2\times(2\pi)^2 {\s^2/\s_0^2}/ (n\times\SNR_\mI^2)$. This indicates that when the $\SNR$ is high enough, this direct estimate of the phase can become better than the estimate coming from an indirect OPD estimator with corresponding phase error variance $[\varPhir(\s)]_\textrm{indirect}$, like the attitude estimator used on BETTII. 

\begin{figure}[ht!]
\begin{center}
\plotone{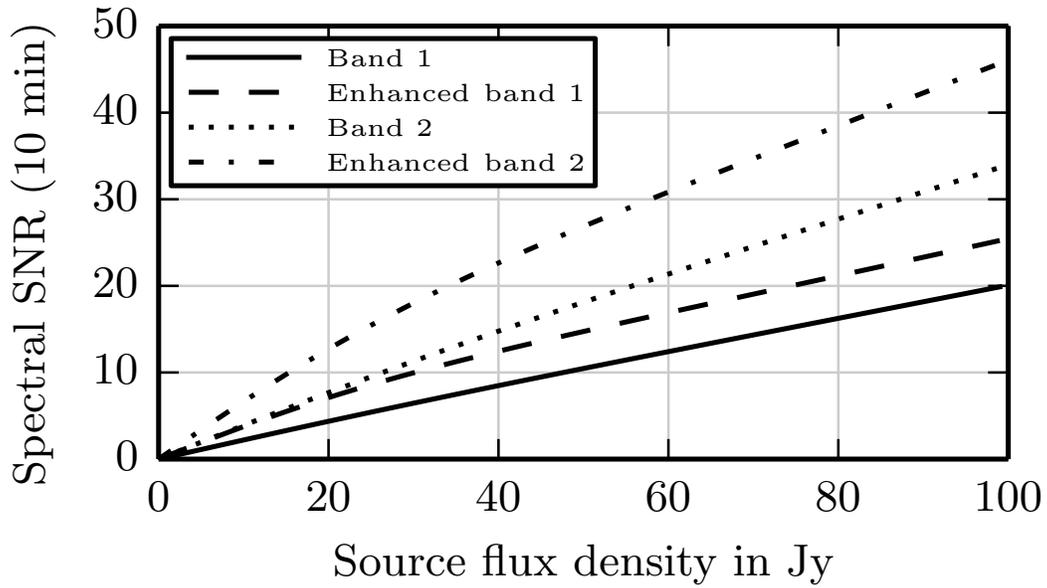}
\caption[BETTII Spectral $\SNR$]{
BETTII's spectral sensitivity. Solid: Normal observing mode, band 1; Dashed: Enhanced sensitivity mode, band 1; Dotted: normal observing mode, band 2; Dot-dashed: Enhanced sensitivity mode, band 2. This plot includes the technique of fringe tracking in the science channel for sufficiently bright sources (see Appendix C). As the source flux rises, the effects of the phase noise become larger and the $\SNR$ should reach an asymptotic value. However, with fringe tracking, the phase noise itself becomes smaller since one can see fringes in one single or a few consecutive scans, so the co-adding becomes easier. Thanks to the fringe-tracking, there is no regime where the phase noise is expected to be dominant on BETTII, provided that the control system performs according to expectations.}
\label{fig:SpectralSNR}
\end{center}
\end{figure}

In Figure~\ref{fig:SpectralSNR}, we use Eq.~\ref{eq:noisephth} and a total phase error variance which is a combination of the phase noise from the direct and indirect methods, to ensure continuity:
\begin{equation}
\varPhir(\s) = \left( \frac{1}{[\varPhir(\s)]_\textrm{direct}} + \frac{1}{[\varPhir(\s)]_\textrm{indirect}}\right)^{-1}.
\end{equation}

On BETTII, the bulk of the phase noise comes from the uncertainties in co-adding consecutive scans (timescale~3), as the estimator uses an indirect method and never really measures the absolute phase for low-$\SNR$ targets. For high-$\SNR$ targets, the method described above can serve as a fringe tracker that not only is useful for calibration, but can also substantially improve the phase estimator's stability over long periods of time by preventing drifts.

\bibliographystyle{chicago}
\bibliography{references.bib}   


\begin{table}[ht!]
\begin{center}
\begin{tabular}{|c|c|c|c|}
\hline 
\textbf{Parameter} & \textbf{Band 1} & \textbf{Band 2} & \textbf{Comment} \\ 
\hline 
Window emissivity & 0.02 & 0.02 & Measured in the lab \\ 
\hline 
Telescope emissivity & 0.077 & 0.077 & 10 mirrors at 0.992 reflection\\ 
\hline 
Sky radiance & 0.16 W.m$^{-2}$.sr$^{-1}$ & 0.07 W.m$^{-2}$.sr$^{-1}$ & \cite{Harries:1980cva} \\ 
\hline 
Window radiance & 0.17 W.m$^{-2}$.sr$^{-1}$ & 0.04 W.m$^{-2}$.sr$^{-1}$ & Blackbody at 240~K \\ 
\hline 
Telescope radiance & 0.17 W.m$^{-2}$.sr$^{-1}$ & 0.04 W.m$^{-2}$.sr$^{-1}$ & Blackbody at 240~K \\ 
\hline 
Total optical efficiency & 0.3 & 0.3 & Per arm, includes detectors \\ 
\hline 
Photon power from the sky & 35 pW & 36 pW & • \\ 
\hline 
Photon power from the window & 40 pW &  18 pW & • \\ 
\hline 
Photon power from the telescope & 108 pW & 38 pW & • \\ 
\hline 
Total absorbed photon NEP & 1.4$\times 10^{-15}$ W.Hz$^{-0.5}$ & 7$\times 10^{-16}$ W.Hz$^{-0.5}$ & From each arm \\ 
\hline 
Detector NEP & 3$\times 10^{-16}$ W.Hz$^{-0.5}$ & 3$\times 10^{-16}$ W.Hz$^{-0.5}$ & For each detector \\ 
\hline 
Total NEP & 2$\times 10^{-15}$ W.Hz$^{-0.5}$ & 1$\times 10^{-15}$ W.Hz$^{-0.5}$ & For the sum of both arms \\ 
\hline 
\end{tabular} 
\caption{BETTII noise parameters}
\label{tab:noise}
\end{center}
\end{table} 
\begin{table}[ht!]
\begin{center}
\begin{tabular}{|c|c|c|c|}
\hline \multicolumn{4}{|c|}{\textbf{Single scan}} \\
\hline 
     & Band 1 &  Band 2 & $\SNR$ Target \\
\hline
MDFD & 15 Jy  & 26 Jy & $\SNR_\mI = 1$\\ 
\hline
\hline \multicolumn{4}{|c|}{\textbf{Normal observing (200 scans, 10 min)}} \\
\hline
MDFD & 1 Jy  & 2 Jy & $\SNR_\mI = 1$\\ 
\hline 
Faintest pt. source & 25 Jy  & 13 Jy & $\SNR_k = 5$\\ 
\hline 
\hline
\multicolumn{4}{|c|}{\textbf{Enhanced sensitivity (200 scans, 10 min)}} \\
\hline
Faintest pt. source & 14 Jy  & 7 Jy & $\SNR_k = 5$\\ 
\hline 
\end{tabular} 
\caption{BETTII sensitivity estimates}
\label{tab:sensitivity}
\end{center}
\end{table}

\end{document}